\documentclass[10pt,final, conference]{IEEEtran}
\usepackage{amssymb}
\usepackage{pstricks}
\usepackage{amsmath}
\usepackage[dvips]{graphicx}
\usepackage{enumerate}
\usepackage{ifthen}
\usepackage{flushend}

\newtheorem{theorem}{Theorem}

\newtheorem{lemma}{Lemma}

\IEEEoverridecommandlockouts

\title{Uplink Performance of Large Optimum-Combining Antenna Arrays in Poisson-Cell Networks}
\author{\IEEEauthorblockN{Siddhartan Govindasamy}
\IEEEauthorblockA{F. W. Olin College of Engineering. Needham, MA,  USA\\
Email: siddhartan.govindasamy@olin.edu}
\thanks{\noindent Sponsored in part by the National Science Foundation under grant CCF-111721.}}

\begin{document}

\maketitle
\pagenumbering{arabic}

\begin{abstract}
The uplink of a wireless network with base stations distributed according to a Poisson Point Process (PPP) is analyzed. The base stations are assumed to have a large number of antennas and use linear minimum-mean-square-error (MMSE) spatial processing for multiple access. The number of active mobiles per cell is limited to permit channel estimation using pilot sequences that are orthogonal in each cell. The cumulative distribution function (CDF) of a randomly located link in a typical cell of such a system is derived when accurate channel estimation is available. A simple bound is provided for the spectral efficiency when channel estimates suffer from pilot contamination. The results provide insight into the performance of so-called massive Multiple-Input-Multiple-Output (MIMO) systems in spatially distributed cellular networks.
\end{abstract}

\begin{keywords}
Massive MIMO, MMSE.
\end{keywords}

\section{Introduction}

Cellular systems with large numbers of base station antennas servicing a relatively small number of mobiles per cell has been proposed as a method to meet the increasing demand for wireless data communications. In such systems, mobiles transmit simultaneously in the same frequency band and the base-station separates the signals from the mobiles spatially \cite{marzetta2010noncooperative, LarssonMag}. As the number of antennas at each base station grows large, the matched-filter (MF) receiver (and its transmit-side analog) are optimal \cite{marzetta2010noncooperative}. However there is a significant range of parameters where the performance of the MMSE receiver greatly exceeds the performance of the simpler MF  receiver \cite{hoydis2013massive} which makes analysis of the MMSE receiver in such networks interesting. The uplink performance of massive MIMO systems with MMSE processing has been analyzed before in \cite{hoydis2013massive} and \cite{YatesMMSEMassive}, but in both those works, the spatial distribution of the network was not explicitly analyzed. Most works which analyze such systems have not explicitly modeled the spatial distribution of base stations and mobiles. Such analyses have the potential to provide valuable insight into the large-scale performance of cellular networks as noted in \cite{AndrewsCellular} which considered the downlink of single-antenna systems in Poisson-cell networks.

In this work, we analyze the performance of the uplink of a spatially distributed cellular system with multi-antenna,  linear MMSE receivers at the base-stations in the interference-limited regime. The base stations are spatially distributed according to a homogenous PPP on the plane.  The mobile nodes are also assumed to be spatially distributed, but unlike in \cite{CellularNetworks} where we assumed that the transmitting nodes are at independent spatial locations, in this work, we limit the number of \emph{active} mobiles per cell to $K$, which results in correlation between the locations of the active mobiles in the network. Limiting the number of active mobiles per cell enables the use of orthogonal pilot sequences for channel estimation in each cell. This is the standard approach for channel estimation assumed in the literature \cite{marzetta2010noncooperative}. Placing a limitation on the number of active mobiles per cell has other practical benefits as well, such as to meet quality of service requirements. However, limiting the number of active mobiles per cell causes the spatial positions of active mobiles in the network to become correlated which significantly complicates analysis. One approach to analyzing networks with spatially correlated users is by making the density of active users small, e.g., through the use of a medium-access-control protocol \cite{GantiHighSIR, giacomelli2011outage}. This approach is not well suited to massive MIMO systems where a central assumption is that multiple mobiles transmit simultaneously in every cell. A second approach, which we proposed in \cite{CorrTransPaper}, is to consider linear MMSE receivers with large numbers of antennas. In this work, we follow  a similar approach, making use of the framework we introduced in \cite{CorrTransPaper} to derive asymptotic expressions for the spectral efficiency (assuming Gaussian codebooks) of a representative link with a large number of receiver antennas as a function of the number of antennas $N$,  link length, mobile and base station density, path-loss exponent and maximum number of active mobiles per-cell, $K$. We also provide the CDF of the spectral efficiency when the representative link is randomly distributed in a typical cell of the network. In addition, we provide bounds to the spectral efficiency for systems where the channel estimate suffers from pilot contamination \cite{jose2011pilot}.

\section{System Model}
Consider a cellular  network with base stations distributed according to a PPP with density $\rho_c$ base stations per unit area and suppose
that the co-ordinates of the system are shifted such that the base-station closest to the origin is shifted to the origin.
Assume that cells are formed by a Voronoi tessellation of the plane with the base stations as generator points. Such a tessellation is referred to as a Poisson-Voronoi Tessellation (PVT).  Let the locations of the base stations be $B_0, B_1, B_2, \cdots$,
with $B_0$  denoting the base station at the origin which we call the representative base station. Suppose that there is a mobile at  $X_0$ which we call the representative
transmitter which is transmitting to the representative base station. In the remainder of this work, we shall analyze  the link between the representative transmitter
and the representative base station which we shall also refer to as the representative link. We shall denote a realization of the base station point process by $\Pi$.
\begin{figure}
\center
\includegraphics[width = 2.75in]{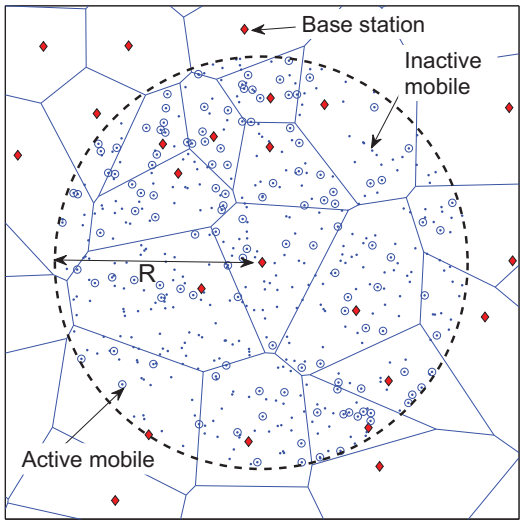}
\caption{Illustration of a Poisson-cell network with maximum number of active mobiles per cell, $K = 10$. Mobiles are represented by the dots and circles are used to highlight the active mobiles.
Observe that smaller cells tend to have a higher density of active mobiles.}
\label{Fig:Cells}
\end{figure}
Overlaid on the network of base stations is a circular network of radius $R$ centered at the origin as shown in Figure \ref{Fig:Cells},
with $n$ additional mobiles independent and identically distributed (i.i.d.) in the circular network. These mobiles, if active will be co-channel interferers to the representative link.
Let the mobiles be located at  $X_1, X_2, \cdots, X_{n}$, and $r_i = |X_i|$ be the distance of the $i$-th mobile from the origin. The area density of mobiles
$\rho_m$ satisfies
\vspace{0.3cm}
\begin{align}
n = \pi\, \rho_m \, R^2. \label{Eqn:MaternPotentialInterferers}
\end{align}

\vspace{0.3cm}

The average power (averaged over the fast-fading random variables) received at each antenna of the representative base station from a mobile at a distance $r_i$, transmitting with power $P_i$ is $P_ir_i^{-\alpha}$, with $\alpha > 2$. The transmit power of the $i$-th mobile, $P_i$ equals one or zero,  depending on whether or not the mobile is active. Hence, we do not consider power control in this work. Up to $K\geq 1$ mobiles are active in each cell.
If a cell has $K$ or less mobiles they are all active, and if a cell has greater than $K$ mobiles, $K$ of the mobiles are selected randomly and with uniform probability to be active.
The representative transmitter is always active and thus $P_0 = 1$. $\mathbf{y}\in\mathbb{C}^{N\times 1}$ contains the sampled signals at a given sampling time at the $N$
antennas of the representative base station and is given by
\begin{align}
\mathbf{y} = r_0^{-\frac{\alpha}{2}} \mathbf{g}_0  x_0 + \sum_{i= 1}^n r_i^{-\frac{\alpha}{2}} \mathbf{g}_i \sqrt{P_i} x_i\,, \label{Eqn:SysEq}
\end{align}
where  $x_i$, is the  transmitted symbol of  the $i$-th mobile and $\mathbf{g}_i \in \mathbb{C}^{N\times 1}$ contains i.i.d., zero-mean,
unit  variance, circularly symmetric, complex Gaussian random variables denoted by $\mathcal{CN}(0,1)$, which represent fast fading between the $i$-th mobile and the $N$ antennas of the representative base station. Since we focus on the interference-limited regime where the noise power $\to$ 0,  \eqref{Eqn:SysEq} does not include noise. Thus, our results are applicable to networks with a high density of mobiles.

We shall consider the limit as $N$, $n$ and $R \to \infty$, such that $n/N = c$ and $\rho_m$ are
constants, and \eqref{Eqn:MaternPotentialInterferers} holds. Note that while our analysis provides
insight into the scaling behavior of such systems, we use it primarily as a tool to analyze large networks of a fixed size.
We additionally require $c > 1/\rho_c$ which  ensures that as $R\to\infty$, with high probability, there would be a
larger number of active mobiles in the \emph{entire} network than antennas at the representative base station. This ensures that the matrix $\mathbf{R}$ defined below is invertible with probability 1 if $N,n,R$ is sufficiently large.

For the rest of this paper,  whenever  $n,N$ or $R \to \infty$, it is assumed that the other two quantities go to infinity as well.
The main results are given in terms of limiting values of a normalized version of
the SIR, $\beta_N = N^{-\alpha/2}r_T^\alpha\mbox{SIR}$, at the output of a linear receiver which estimates $x_0$ using a weight vector
$\mathbf{w}$. The estimate, $\hat{x}_0 = \mathbf{w}^\dagger \mathbf{y}$, and the weight vector,
\begin{align}
&\mathbf{w}^\dagger = \hat{\mathbf{h}}^\dagger\left(\sum_{j = 1}^n r_j^{-\alpha}{P_j} \mathbf{g}_{j} \mathbf{g}_j^\dagger\right)^{-1}\,, \label{Eqn:PCWeight}
\end{align}
where $\hat{\mathbf{h}}$ is an estimate of the channel  vector $\mathbf{g}_0$. Note that the weight vector above is the minimum-mean-square-error weight vector if $\hat{\mathbf{h}}= \mathbf{g}_0$
which is the assumption we use for the first part of this work. We also consider a pilot-contaminated estimate of $\mathbf{g}_0$  in Section \ref{Sec:PilotCont}.
We assume that the interference covariance matrix
\begin{align}
\mathbf{R} = \sum_{j = 1}^n r_j^{-\alpha}{P_j} \mathbf{g}_{j} \mathbf{g}_j^\dagger
\end{align}
is known perfectly at the representative base station. 
The SIR at the output of the receiver is given by
\begin{align}
  \text{SIR} = \frac{r_0^{-\alpha}\left|\mathbf{w}^\dagger \mathbf{g}_0\right|^2}{\sum_{i = 1}^n r_i^{-\alpha}P_i\left|\mathbf{w}^\dagger \mathbf{g}_i\right|^2} \label{Eqn:GenSIR}
\end{align}

\section{Main Results}
\subsection{Density of Active Transmissions}
 The density of active mobiles in our system  is the product of the density of all mobiles and the probability that a mobile is active in the limit as $R\to\infty$. This quantity is used in characterizing the spectral efficiency in the subsequent sections and is interesting in its own right as it determines the fraction of mobiles that can be active at any one time as a function of $K$, $\rho_c$ and $\rho_m$. The probability that a mobile is active  is dependent the distribution of cell sizes, and is given in the following lemma.
\begin{lemma}\label{Sec:ActiveNodeDensityLemma}
\begin{align}
\lim_{n\to\infty}\Pr(P_j =1) = \rho_c\,E\left[|\mathcal{C}^T|h(|\mathcal{C}^T|)\right]\,.\
\end{align}
where the expectation is with respect to the PDF of the area of the \emph{typical cell} of a PVT with density $\rho_c$, and
\begin{align}
&h(a)\triangleq\frac{K-Ke^{-\rho_m a}\sum_{k = 0}^K \!\frac{(\rho_m a)^k}{k!}}{\rho_m a} +\sum_{m= 0}^{K-1}\frac{\left({\rho_m a}\right)^m}{m!}\!  \!e^{-\rho_m a}. \label{Eqn:InSumPowerTerm} \end{align}
\end{lemma}
{\it Proof:} Please see in  Appendix \ref{Sec:ActiveNodeDensityLemmaProof}.

The typical cell (see e.g. \cite{kendall2010new} for its precise definition)  is statistically equivalent to the cell containing the origin if the origin is added to the set of generator points of a PPP that underlies a PVT (see e.g. \cite{kendall2010new}). The exact PDF of $|\mathcal{C}^T|$ is given in \cite{calka2003precise}, but it is expressed as an infinite series involving multiple integrals, which is challenging to compute numerically (e.g. the authors use monte-carlo integration to evaluate it in \cite{calka2003precise}).  We do not include the explicit expressions here for the sake of brevity. Instead, we use an approximation to the PDF of $a =|\mathcal{C}^T|$, which is given in terms of a generalized gamma PDF as follows \cite{tanemura2003statistical}.
For $a > 0$,
\begin{align}
f_A(a) \approx {15.225\,\rho_c\, (\rho_c\,a)^{2.311} e^{-3.032\, (\rho_c\,a)^{1.080}}}
\end{align}

\subsection{Spectral Efficiency with Perfect Channel Estimation}
Assume that the representative base station has a perfect estimate of the channel vector, i.e. $\hat{\mathbf{h}}=\mathbf{g}_0$. The normalized SIR at the output of the MMSE receiver is then
\begin{align}
\beta_N  = N^{-\alpha/2}\hat{\mathbf{h}}^\dagger \mathbf{R}^{-1} \hat{\mathbf{h}}\,.
\end{align}

Conditioned on a specific realization of the base-station point process $\Pi$ and the length of the representative link $r_0$, which is assumed to be active, we have the following theorem.
\begin{theorem}
If the system model from the previous section holds, as $n, N, R\to\infty$ such that $n/N = c > 0$, and \eqref{Eqn:MaternPotentialInterferers} hold, $\beta_N \to \beta$ in probability where $\beta$ is the real, positive
solution to
\begin{align}
\frac{2\pi^2\rho \beta^{\frac{2}{\alpha}} }{\alpha}&\csc\left(\frac{2\pi}{\alpha}\right)  =  1 + \frac{2(\pi \rho)^{2-\frac{2}{\alpha}}\beta}{(\alpha -2)(c + \pi \rho \beta)^{1-\frac{2}{\alpha}}}\; \times \nonumber \\
&_2F_1\left(1-\frac{2}{\alpha}, 1-\frac{2}{\alpha}; 2-\frac{2}{\alpha}; \frac{\pi \rho \beta}{ \pi \rho \beta + c}\right)\,, \label{Eqn:LimitingTxCSISINRThinned}
\end{align}
where $\rho = \rho_m \lim_{n\to\infty}\Pr(P_j=1)$, and $_2 F_1(.,.;.;.)$ is the Gauss hypergeometric function.
\end{theorem}
{\it Proof:} Given in Appendix \ref{Sec:MainProof}.

\noindent Note that $\rho$ is  the density of mobiles in the limit.  Additionally, when the number of users in the \emph{entire} network is much larger than the number of antennas at the representative base station, i.e. $n  \gg N$, the second
term on the RHS of \eqref{Eqn:LimitingTxCSISINRThinned} is small \cite{CorrTransPaper}. Approximating that term by zero, we can find an approximation for $\beta$ which when combined with the Shannon equation, yields the following expression for the spectral efficiency and its mean when Gaussian codebooks are used by each mobile \cite{CorrTransPaper}.
\begin{align}
\gamma\approx E[\gamma]\approx \log_2\left(1+\left[\frac{N\,\alpha}{2\pi^2\rho r_0^2}\sin\left(\frac{2\pi}{\alpha}\right)\right]^{\frac{\alpha}{2}}\right). \label{Eqn:MeanSpecEffApprox}
\end{align}

If the representative transmitter is distributed with uniform probability in the cell at the origin, we can derive the CDF of the spectral efficiency from \eqref{Eqn:MeanSpecEffApprox} using the nearest-neighbor distribution of a PPP (e.g. see \cite{kendall2010new}). This yields the following approximation to the CDF of the spectral efficiency of the representative link assuming a large number of base station antennas $N$ and the number of users in the network greatly exceeding the number of antennas at the base station $n \gg N$.
\begin{align}
\Pr(\gamma \leq \tau) \approx e^{-\frac{\rho_c}{\rho}N\frac{\alpha}{2\pi}\sin\left(\frac{2\pi}{\alpha}\right)\left(\frac{1}{2^\tau - 1}\right)^{\frac{2}{\alpha}}}\,. \label{Eqn:CDFSpecEff}
\end{align}

\subsection{Pilot Contaminated Channel Estimation}\label{Sec:PilotCont}

We assume that the pilot signals used for channel estimation by the mobiles in a given cell are orthogonal, but that the same set of pilot signals is repeated in different cells. The estimated channel vector of the representative transmitter suffers from pilot contamination \cite{jose2011pilot} from mobiles in other cells who shared the same pilot sequence. If we assume that the power used during the training sequence is high compared to the noise, we can neglect the effect of noise in the channel estimate. Thus the estimated channel between the representative base station and the representative transmitter $\hat{\mathbf{h}}$ is given by,
\begin{align}
\hat{\mathbf{h}} = \sum_{i\in\mathcal{T}} r_i^{-\frac{\alpha}{2}}\sqrt{P_i} \mathbf{g}_{i} \label{Eqn:PilotContaminatedChannel}
\end{align}
where $\mathcal{T}$ is the set of indices of the mobiles (including the representative transmitter) which shared the same pilot sequence with the representative transmitter during channel estimation. For the purposes of this section, we assume that $\alpha > 4$ which enables us to simplify the effect of the pilot contamination.

We assume that the receiver uses the weight vector in \eqref{Eqn:PCWeight} with the pilot-contaminated channel estimate $\hat{\mathbf{h}}$ from \eqref{Eqn:PilotContaminatedChannel}. Note that we continue to assume that the interference covariance matrix is known at the receiver. We refer to this receiver as the PC-MMSE receiver. As in the proof of Theorem 1, we characterize the normalized SIR for the PC-MMSE receiver for a fixed $r_0$. The normalized SIR for the PC-MMSE receiver is denoted by $\bar{\beta}_N=N^{-\alpha/2} r_0^\alpha$ SIR. We can now state the following theorem.
\begin{theorem} \label{Theorem:PCSIR}
Conditioned $\Pi$ and $r_0$, $\bar{\beta}_N\to\bar{\beta}$  in probability, with $\bar{\beta}$ is bounded from below by the following random variable
\begin{align}
\bar{\beta}_N &\to \bar{\beta} \geq\frac{r_0^{-2\alpha}}{r_0^{-\alpha} + \sum_{\substack{j\in\mathcal{T}\\ j\neq0}}  {r_j^{-\alpha}{P_j}}}\beta \end{align}
\end{theorem}
{\it Proof:} Given in Appendix \ref{SEC:TheoremPCSIRProof}.

We can further bound $\bar{\beta}$ by assuming that there is exactly one mobile in every cell that shares the pilot sequence with the representative transmitter during channel estimation. Furthermore, for $j > 0$, observe that the closest point to the origin in the cell  associated with the base station at $B_j$ is bounded from above by $|B_j|/2$. Thus, all mobiles located in the cell associated with the base-station at $B_j$ are at a distance $B_j/2$ or greater from the base station at the origin. This leads to the following bound.
\begin{align}
\bar{\beta}\geq\frac{r_0^{-2\alpha}}{r_0^{-\alpha} + \sum_{j = 1}^\infty  {\left|\frac{B_j}{2}\right|^{-\alpha}}}\beta \label{Eqn:PCBound}
\end{align}
Note here that while the bound given above is loose, to the best of our knowledge it is the only such bound which considers the effect of spatially distributed pilot contaminators.
\section{Numerical Simulations and Results}
\begin{figure}
\center
\includegraphics[width = 3.5in]{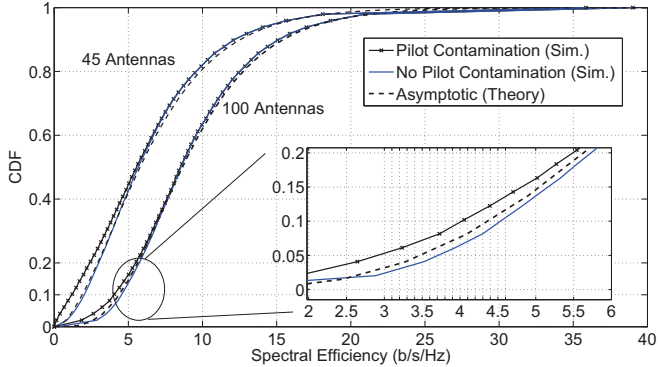}
\caption{Simulated CDF of the spectral efficiency for 45 and 100 antennas and with and without pilot contamination. The parameters used are $\rho_c = 2\times 10^{-5}$, $\rho_m  = 0.001, \alpha =  5$ and the representative transmitter was randomly placed in the cell at the origin. The maximum number of active mobiles per cell is $K=10$.}
\label{Fig:CDFPC}
\end{figure}

We conducted Monte Carlo simulations to verify the accuracy of our asymptotic results. Figure \ref{Fig:CDFPC} shows the CDFs of the spectral efficiency
from simulations with and without pilot contamination as well as the asymptotic expression for the CDF from \eqref{Eqn:CDFSpecEff} for 45 and 100 antennas at the base station. The representative link was randomly distributed in the cell containing the origin. The remaining parameters are given in the figure caption. Observe that with 100 antennas and at an outage probability of 0.1, the theoretical prediction is within 0.25 b/s/Hz of the simulated values which validates \eqref{Eqn:CDFSpecEff}. For 45 antennas, the asymptotic CDF is within 0.3 b/s/Hz of the simulated values. Additionally, note that the simulated pilot contaminated spectral efficiency for 100 antennas is within 0.5 b/s/Hz of the non-pilot contaminated spectral efficiency.

\begin{figure}
\center
\includegraphics[width = 3.5in]{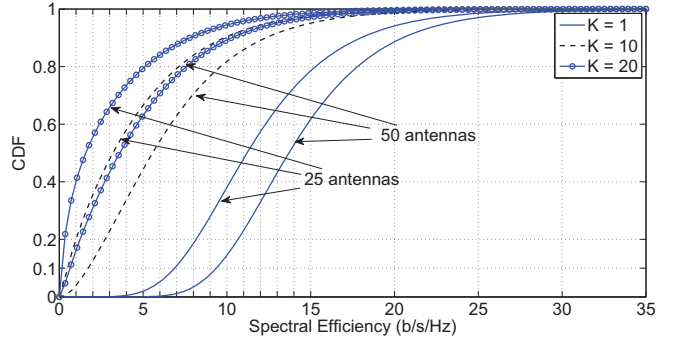}
\caption{Simulated CDF of spectral efficiency for 25 and 50 antennas. The parameters used are $\rho_c = 2\times 10^{-5}$, $\rho_m  = 0.001, \alpha =  5$ and the representative transmitter was randomly placed in the cell at the origin. The maximum number of active mobiles per cell is $K=1, 10$ and $20$.
}
\label{Fig:CDFComp}
\end{figure}

In Figure \ref{Fig:CDFComp}, we plotted the CDFs from \eqref{Eqn:CDFSpecEff} for 25 and 50 antennas per base station and $K = 1, 10$ and $20$. This figure illustrates how the results in this paper can be used to  analyze the tradeoff between increasing the density of transmissions by increasing $K$ and the resulting reduction in per-link data rates due to increased interference.  For a system with 50 antennas, at an outage probability of 0.1, there is approximately a five-fold increase in the spectral efficiency going from $K = 10$, to $K = 1$. The density of active mobiles with $K = 10$ is approximately $3.8\times 10^{-4}$ and the density of active mobiles with $K = 1$ is approximately $4\times 10^{-5}$. Thus an approximately 10 fold increase in mobile density  results in a reduction in the spectral efficiency of the representative link by approximately a factor of five, at an outage probability of 0.1. On the other hand, going from $K = 20$ to $K= 10$ results in nearly a doubling of the spectral efficiency of a representative link but with the density of active transmissions reduced by approximately a factor of less than two. Combined with models for channel coherence and training times (which depend on $K$), such an analysis could be used to optimize $K$.

\begin{figure}
\center
\includegraphics[width = 3.5in]{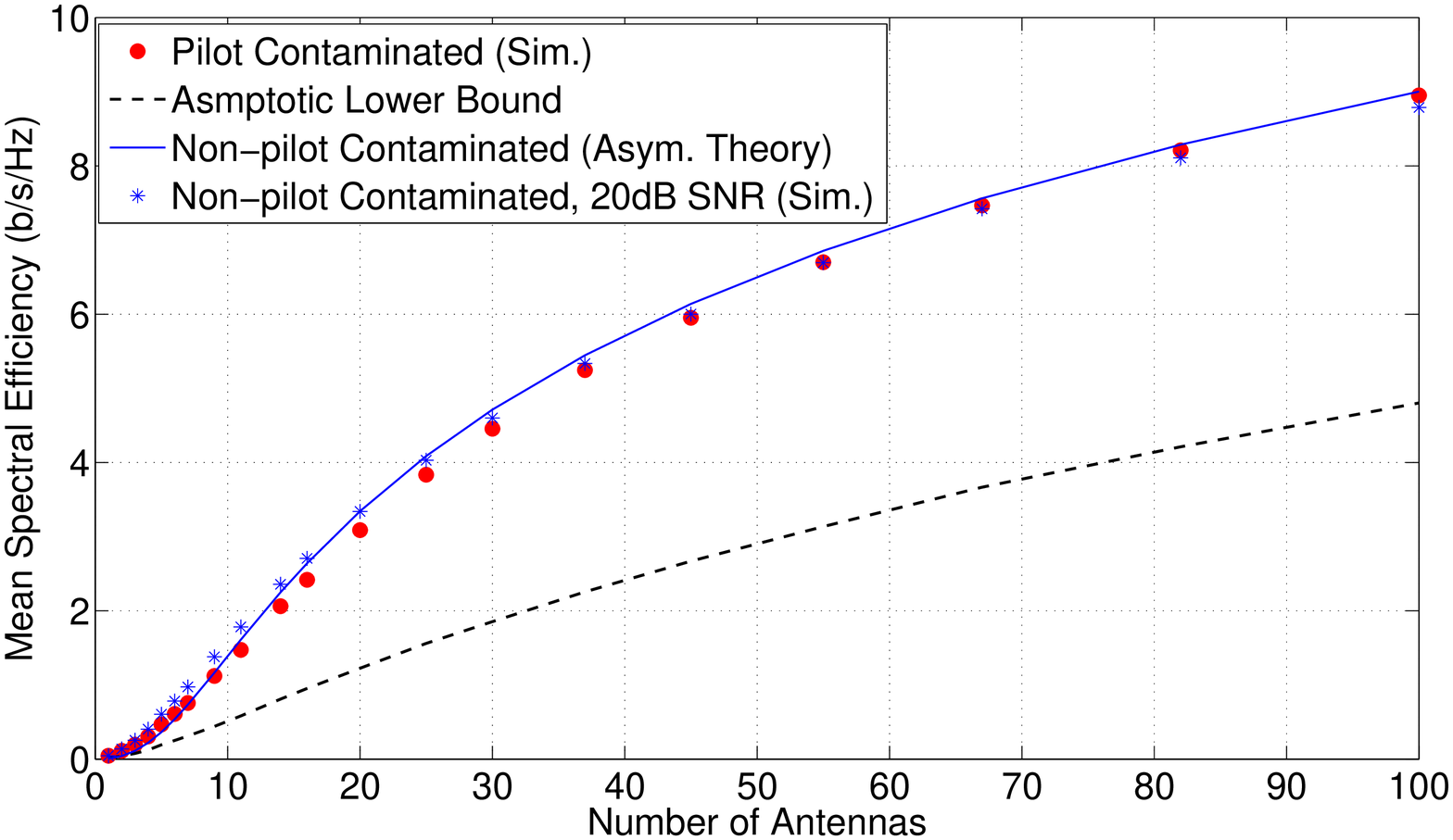}
\caption{Mean spectral efficiency vs. number of base station antennas with pilot contamination for a system with $\rho_c = 0.00002, \rho_m = 0.001, K = 10$, $\alpha = 5$,  and a fixed link length $r_T = 100$. The dashed lines represent a lower bound for the mean spectral efficiency. 
}\label{Fig:PCBound}
\end{figure}

Figure \ref{Fig:PCBound} illustrates the lower bound for the spectral efficiency for a fixed link length of $r_T = 100$ and $K = 5$, and the remaining parameters as given in the caption. The simulated markers represent the pilot contaminated mean spectral efficiency and the dashed line represents the lower bound from \eqref{Eqn:PCBound}. For reference, the simulated asymptotic spectral efficiency with perfect channel estimation plotted using the solid line, and simulated spectral efficiencies for a system with non-zero noise such that the Signal-to-Noise-Ratio (SNR) is 20dB is plotted using the asterisk markers. Note that the lower bound is loose. However, it provides a guarantee on the worst case mean spectral efficiency under pilot contamination. Additionally, the simulations with 20dB SNR indicate that the interference-limited approximation is accurate even when the number of antennas is large.

\section{Summary and Conclusions}
In this paper, the spectral efficiency of the uplink in a spatially distributed, Poisson-cell network is analyzed. The base station is assumed to use a linear MMSE estimator and the number of active mobiles per cell is limited to $K$ in order to permit channel estimation using intra-cell orthogonal pilots.   The CDF of the spectral efficiency in the interference-limited regime is derived for a randomly distributed mobile in a typical cell of the network. The results can be used to statistically characterize achievable data rates in such networks as a function of tangible system parameters such as user and base-station density and number of antennas, and can help system designers optimize parameters such as the maximum  number of active mobiles per cell.

\appendix
\subsection{Proof of Main Result} \label{Sec:MainProof}
The main result is proved using Theorem 1 of \cite{CorrTransPaper} in which the normalized SIR of a representative link with $N$ antennas and a linear MMSE receiver in a network with interferers at correlated spatial positions is shown to converge in probability to a limit of the same form as in Theorem 1 in this paper. Theorem 1 in this paper follows directly from that result if the following two conditions hold.
\begin{align}
&\lim_{n\to\infty} \Pr(P_i r_i^{-\alpha} N^{\frac{\alpha}{2}}\leq x, P_j r_j^{-\alpha}N^{\frac{\alpha}{2}} \leq x) \notag \\
&\;\;\;\;\;= \lim_{n\to\infty}\Pr(P_i r_i^{-\alpha}N^{\frac{\alpha}{2}} \leq x)\Pr(P_j r_j^{-\alpha}N^{\frac{\alpha}{2}} \leq x)\,. \label{Eqn:AsympIndepSym}
\end{align}
and
\begin{align}
\lim_{n\to\infty} \Pr(r_i^{-\alpha} N^{\frac{\alpha}{2}} \leq x | P_i = 1) = \lim_{n\to\infty}\Pr(r_i^{-\alpha} N^{\frac{\alpha}{2}} \leq x)\,. \label{Eqn:DistanceFactor}
\end{align}
Equation  \eqref{Eqn:AsympIndepSym} is shown in  Appendix \ref{Sec:PowerFactorLemmaProof}, and \eqref{Eqn:DistanceFactor} follows directly from the following lemma
\begin{lemma} \label{Lemma:PowerPathLossFactor}
For any $x>0$,
\begin{align}
\lim_{N\to\infty}\Pr\left(P_i  = 0| r_i \leq x^{-1/\alpha} \sqrt{N}\right)= \lim_{N\to\infty}\Pr\left(P_i  = 0\right)\notag
\end{align}
\end{lemma}
{\it Proof: }The proof follows similar steps to that  used to prove Lemma \ref{Sec:ActiveNodeDensityLemma} and is omitted for brevity.

\subsection{Proof of Lemma \ref{Sec:ActiveNodeDensityLemma}}\label{Sec:ActiveNodeDensityLemmaProof}

Let $C_i$ denote the cell occupied by $X_i$. Let $A_i = \left|C_i \cap B(0,R)\right|$ be the area of intersection between $C_i$ and  $B(0,R)$. Additionally let $\#C_i$ denote the number of mobiles in $C_i$ \emph{excluding} $X_i$, and $\mathcal{A}$ be the event that $X_i$ and $X_j$ are in diferrent cells, i.e.
$C_i \neq C_j$. Since the cells of a PVT are finite with probability 1 (e.g. see \cite{kendall2010new}),  as $R\to\infty$, $\Pr(\mathcal{A}) \to 1$. Recall that all mobiles in a given cell are active if there are $K$ or fewer mobiles in the cell.
If there are greater than $K$ mobiles, $K$ are selected randomly to be active. Thus,
\begin{align}
\Pr(P_j =1|\#C(X_j) = \ell, \mathcal{A}) = \begin{cases} 1, & \mbox{if } \ell < K \\ \frac{K}{\ell+1}, & \mbox{otherwise. } \end{cases} \label{Eqn:ProbActiveGivenNums}
\end{align}
For $n-2 \geq m$, we have
\begin{align}
&\Pr(\#C(X_j) = m| A_i = a_i,A_j = a_j, \mathcal{A})  \notag\\
&\;\;\;\;\;\;\;\;\;\;\;\;=\Pr(\#C(X_j) = m| A_j = a_j, \mathcal{A})  \notag\\
&\;\;\;\;\;\;\;\;\;\;\;\;=  {n-2\choose m} \left(\frac{a_j}{\pi R^2}\right)^m \left(\frac{\pi R^2-a_j}{\pi R^2}\right)^{n-m-2}\,. \label{Eqn:OneCellNumNodesConditionFinite}
\end{align}

Let $\mathcal{C}_1, \mathcal{C}_2\cdots$ denote the cells in the network. Define the following set which consists of the portions of the cells wholly contained in  $B(0,d)$.
\begin{align}
\mathcal{C}_d := \{ \mathcal{C}_i \cap B(0,d) : i = 1, 2,\cdots \}\,.
\end{align}
The $j$-th element of $\mathcal{C}_d$ is denoted by $\mathcal{C}_{dj}$. Thus, $A_j$ takes values in the set $\mathcal{C}_{R}$, where  $\Pr(A_i = \mathcal{C}_{Rk}) = |\mathcal{C}_{Rk}|/(\pi R^2)$. Combining \eqref{Eqn:MaternPotentialInterferers}, \eqref{Eqn:ProbActiveGivenNums} and \eqref{Eqn:OneCellNumNodesConditionFinite}, weighting by $\Pr(A_i = \mathcal{C}_{Rk})$  and summing over all possible values of $A_j$ yields,
\begin{align}
&\Pr(P_j = 1|\mathcal{A})
=\sum_{k = 1}^{|\mathcal{C}_R|}\frac{|\mathcal{C}_{Rk}|}{\pi R^2}\left[\sum_{\ell = 0}^{n-2}\text{min}\left(1,\frac{K}{\ell+1}\right) \right.\times  \notag\\
&\left.{n-2\choose \ell} \left(\frac{\rho_m\,|\mathcal{C}_{Rk}|}{n}\right)^\ell\left(1-\frac{\rho_m\,|\mathcal{C}_{Rk}|}{n}\right)^{n-\ell-2} \right] \label{Eqn:SinglePowerLimit}
\end{align}
The next step is  to take the limit of \eqref{Eqn:SinglePowerLimit} as $n, R\to\infty$.
Since the term in the brackets is bounded for all $n$,
we can its limit before taking the outer limit in \eqref{Eqn:SinglePowerLimit} (see e.g. \cite{Habil}).
Writing $a = |\mathcal{C}_{Rk}|$, the limit of the term in the brackets in \eqref{Eqn:SinglePowerLimit} is
\begin{align}
&\lim_{n\to\infty}\left[\sum_{\ell = K}^{n-2}\frac{K}{\ell+1} {n-2\choose \ell} \left(\frac{\rho_m\,a}{n}\right)^\ell\left(1-\frac{\rho_m\,a}{n}\right)^{n-\ell-2}{a} \right.\notag\\
&\left. + \sum_{m= 0}^{K-1}{n-2\choose m} \left(\frac{\rho_m\,a}{n}\right)^m\!\!\left(1-\frac{\rho_m\,a}{n}\right)^{n-m-2}\!\right] \notag\\
&=\frac{K-Ke^{-\rho_m a}\sum_{k = 0}^K \frac{(\rho_m a)^k}{k!}}{\rho_m a} +\sum_{m= 0}^{K-1}\frac{1}{m!} \left({\rho_m a}\right)^m e^{-\rho_m a}  \notag\\
&= h(a)\,.\label{Eqn:InSumPowerTerm} \end{align}
Thus, we can write the limit of  \eqref{Eqn:SinglePowerLimit} as
\begin{align}
\lim_{R\to\infty}\Pr(P_j = 1|\mathcal{A}) =\lim_{R\to\infty} \frac{|\mathcal{C}_R|}{\pi R^2}\sum_{k = 1}^{|\mathcal{C}_R|}\frac{|\mathcal{C}_{Rk}|}{|\mathcal{C}_R|} h\left(|\mathcal{C}_{Rk}|\right). \label{Eqn:PowLimH}
\end{align}

We next state a result from stochastic geometry (Equation 5.2 of \cite{kendall2010new}), which relates the spatial average of a function over the cells of a single realization of a PVT to the ensemble average. For any bounded function $f$ which maps convex sets in $\mathbb{R}^2$ to $\mathbb{R}$, the following holds with probability 1,
\begin{align}
\lim_{R\to\infty} \frac{1}{|\mathcal{C}_R|} \sum_{k = 1}^{|\mathcal{C}_R|} f(\mathcal{C}_{Rk}) = \frac{1} {E\left[\frac{1}{|\mathcal{C}^0|}\right]}E\left[\frac{f(\mathcal{C}^0)}{|\mathcal{C}^0|}\right]\,,\label{Eqn:ErgodicTypicalCell}
\end{align}
where $\mathcal{C}^0$ is  the \emph{zero-cell} of a PVT, which is statistically equivalent to the cell that contains any given point in the plane.
Additionally, we note that with probability 1, as $R\to\infty$, the number of cells per unit area in $B(0,R)$ approaches
the density of the base stations with probability 1, i.e. $\lim_{R\to\infty}\frac{|\mathcal{C}_{R}|}{\pi R^2} = \rho_c$.
Let $f(\mathcal{C}_{Rk}) = |\mathcal{C}_{Rk}| h(|\mathcal{C}_{Rk}|)$, which is only dependent on the area of $\mathcal{C}_{Rk}$. Thus, we can apply \eqref{Eqn:ErgodicTypicalCell} cellular model we assume here, where the coordinates are shifted such that there is a base station at the origin.
Combining \eqref{Eqn:ErgodicTypicalCell}, the fact that $\Pr(\mathcal{A})\to 1$  and $E\left[\frac{1}{|\mathcal{C}^0}|\right]= \rho_c$ with \eqref{Eqn:PowLimH} we have,
\begin{align}
\lim_{n\to\infty}\Pr(P_j =1) = E\left[{h(|\mathcal{C}^0|)}\right]\,.
\end{align}

Next we express the equation above in terms of the \emph{typical cell} of a PVT which we denote by $\mathcal{C}^T$.
The relationship between the mean of functions of the zero-cell and the typical cell is given by Equation 2.1 of \cite{mecke1999relationship}. Applying this result to the previous equation yields
\begin{align}
\lim_{n\to\infty}\Pr(P_j = 1) = \frac{1}{E[|\mathcal{C}^T|]}E\left[|\mathcal{C}^T|h(|\mathcal{C}^T|)\right]\,.
\end{align}
Substituting the fact that $E[|\mathcal{C}^T|] = \frac{1}{\rho_c}$ with the above expression proves the lemma.

\subsection{Proof of Equation \eqref{Eqn:AsympIndepSym}}\label{Sec:PowerFactorLemmaProof}

Recalling the definitions used in the proof of Lemma \ref{Sec:ActiveNodeDensityLemma}, and following steps similar to that used to prove it, we can write
\begin{align}
&\Pr\left(P_i = 1, P_j = 1 | A_i =|\mathcal{C}_{Rs}|,A_j =|\mathcal{C}_{Rt}|,\mathcal{A} \right) = \notag\\
&\sum_{\ell=0}^{n-2} \text{min}\left(1,\frac{K}{\ell+1}\right){n-2 \choose \ell}\left( \frac{\rho_m|\mathcal{C}_{Rt}|}{n} \right)^\ell \nonumber \\
&\;\;\;\;\;\;\;\;\;\;\;\;\;\;\;\times \left( 1-\frac{\rho_m|\mathcal{C}_{Rt}|}{n} \right)^{n-\ell-2}\times\notag\\
&\sum_{m = 0}^{n-\ell-2}\text{min}\left(1,\frac{K}{m+1}\right){n-\ell-2 \choose m}\left(\frac{\rho_m|\mathcal{C}_{Rs}|}{n- \rho_m|\mathcal{C}_{Rt}|} \right)^m\notag\\
 &\;\;\;\;\;\;\;\;\;\;\;\;\;\;\;\times\left( 1-\frac{\rho_m|\mathcal{C}_{Rt}|}{n- \rho_m|\mathcal{C}_{Rt}|} \right)^{n-m-\ell-2} \label{Eqn:TwoPowers}
\end{align}
Applying steps similar to those used to prove Lemma \ref{Sec:ActiveNodeDensityLemma} to \eqref{Eqn:TwoPowers}, i.e. by weighting \eqref{Eqn:TwoPowers} by the probabilities that $A_i =|\mathcal{C}_{Rs}|,A_j =|\mathcal{C}_{Rt}|$, summing over all $\mathcal{C}_{Rs}$ and $\mathcal{C}_{Rt}$, taking the limit as $n,N, R\to\infty$, and applying \eqref{Eqn:ErgodicTypicalCell}, we can show that
\begin{align}
&\lim_{R\to\infty}\Pr\left(P_i = 1, P_j = 1 | \mathcal{A} \right) = \notag\\
&\;\;\;\;\;\;=\rho_c\,E\left[|\mathcal{C}^T|h(|\mathcal{C}^T|)\right]\rho_c\,E\left[|\mathcal{C}^T|h(|\mathcal{C}^T|)\right]\,. \nonumber \\
&\;\;\;\;\;\;= \lim_{R\to\infty}\Pr\left(P_i = 1\right) \lim_{R\to\infty}\Pr\left(P_j = 1\right)
\end{align}
Since $x > 0$ and $\Pr(\mathcal{A})\to 1$, we have
\begin{align}
&\lim_{R\to\infty}\Pr\left(P_i N^{\alpha/2} r_i^{-\alpha}> x, P_j N^{\alpha/2} r_j^{-\alpha}> x \right)\notag\\
& = \lim_{R\to\infty}\Pr\left(N^{\alpha/2} r_i^{-\alpha}> x, N^{\alpha/2} r_j^{-\alpha}> x | \right. \notag\\
&\;\;\;\;\;\;\;\;\;\; \left. P_i = 1, P_j = 1, \mathcal{A}\right)\Pr\left(P_i = 1, P_j = 1 | \mathcal{A} \right)\notag
\end{align}
Combining this with the fact that $r_i$ and $r_j$ are asymptotically independent when conditioned on $\mathcal{A}$ and $\Pr(P_i = 1, P_j = 1)$,
\begin{align}
&\lim_{R\to\infty}\Pr\left(P_i N^{\alpha/2} r_i^{-\alpha}> x, P_j N^{\alpha/2} r_j^{-\alpha}> x \right)\notag\\
&\;\;\;\;\;\;\;\;\;= \lim_{R\to\infty}\Pr\left(P_i = 1\right)\lim_{R\to\infty}\Pr( N^{\alpha/2} r_i^{-\alpha}> x )\nonumber\\
&\;\;\;\;\;\;\;\;\;\times \lim_{R\to\infty}\Pr\left(P_j = 1\right)\lim_{R\to\infty}\Pr\left(N^{\alpha/2} r_j^{-\alpha}> x \right)
\end{align}
Equation \eqref{Eqn:AsympIndepSym}  follows from Lemma \ref{Lemma:PowerPathLossFactor} and the fact that for $x > 0$, $\Pr\left(P_j N^{\alpha/2} r_j^{-\alpha}> x| P_j = 0 \right) = 0$.

\subsection{Proof of Theorem \ref{Theorem:PCSIR}} \label{SEC:TheoremPCSIRProof}

  Note that $R, n,$ and $N$ are related such that the CDF of $N^{\alpha/2}r_i^{-\alpha/2}$ does not vary with $N,n, R$ and moreover, $p_i=  N^{\alpha/2}r_i^{-\alpha}$ are  bounded for all $n, N, R$. The signal power normalized by $N^{-\alpha}$  at the output of the  PC-MMSE filter is
\begin{align}
&N^{-\alpha}S = r_0^{-\alpha}  N^{-\alpha}  \left| \mathbf{w}^\dagger\mathbf{g}_0\right|^2\notag\\
&=r_0^{-\alpha}\left|\sum_{i \in\mathcal{T}}r_i^{-\frac{\alpha}{2}} \sqrt{P_i}\frac1N\mathbf{g}_i^\dagger\left(\sum_{j = 1}^n\frac1Np_j\mathbf{g}_{j} \mathbf{g}_j^\dagger\right)^{-1}\mathbf{g}_0\right|^2\notag\\
&=r_0^{-\alpha}\left|\sum_{\substack{i \in\mathcal{T},  i\neq 0}}r_i^{-\frac{\alpha}{2}} \sqrt{P_i}\frac1N\mathbf{g}_i^\dagger\left(\sum_{j = 1}^n \frac1N p_i\mathbf{g}_{j} \mathbf{g}_j^\dagger\right)^{-1}\!\!\!\!\mathbf{g}_0\right. \nonumber\\
&\left.+{r_0^{-{\alpha/2}} \frac1N\mathbf{g}_0^\dagger\left(\sum_{j = 1}^n \frac1N p_j \mathbf{g}_{j} \mathbf{g}_j^\dagger\right)^{-1}\mathbf{g}_0}\right|^2\,. \label{Eqn:SINRExpand}
\end{align}
Consider the expectation of the first term in the absolute value in $\eqref{Eqn:SINRExpand}$.
\begin{align}
&E\left[\left|\sum_{\substack{i \in\mathcal{T},  i\neq 0}}r_i^{-\frac{\alpha}{2}} \sqrt{P_i}\frac1N\mathbf{g}_i^\dagger\left(\sum_{j = 1}^n \frac1N p_i\mathbf{g}_{j} \mathbf{g}_j^\dagger\right)^{-1}\!\!\!\!\mathbf{g}_0\right|\right]\leq \notag \\
&E\left[\sum_{\substack{i \in\mathcal{T}, i\neq 0}}\left|r_i^{-\frac{\alpha}{2}} \sqrt{P_i}\right|\left|\frac1N\mathbf{g}_i^\dagger\left(\sum_{\substack{j = 1 ,  j\neq i}}^n \frac1N p_i\mathbf{g}_{j} \mathbf{g}_j^\dagger\right)^{-1}\!\!\!\!\mathbf{g}_0\right|\right]\,,\notag
\end{align}
where the inequality follows from applying the Sherman-Morrison-Woodbury matrix inversion lemma and the non-negative definiteness of the matrix
\begin{align}
\left(\sum_{\substack{j = 1, j\neq i}}^n \frac1N p_i\mathbf{g}_{j} \mathbf{g}_j^\dagger\right)\,.
\end{align}
From Lemma 3 of \cite{CorrTransPaper} if \eqref{Eqn:AsympIndepSym} and \eqref{Eqn:DistanceFactor} hold, $\exists n_0$, such that for alll $n > n_0$, with probability 1, the minimum eigenvalue of the matrix above is bounded from below by $\lambda_{\ell b} > 0$. For $n > n_0$, taking an eigen-decomposition of this matrix and simplifying:
\begin{align}
&E\left[\left|\sum_{\substack{i \in\mathcal{T}, i\neq 0}}r_i^{-\frac{\alpha}{2}} \sqrt{P_i}\frac1N\mathbf{g}_i^\dagger\left(\sum_{\substack{j = 1,j\neq i}}^n \frac1N p_i\mathbf{g}_{j} \mathbf{g}_j^\dagger\right)^{-1}\!\!\!\!\mathbf{g}_0\right|\right]\notag\\
&\leq E\left[\left|\sum_{\substack{i \in\mathcal{T}, i\neq 0}}r_i^{-\frac{\alpha}{2}} \sqrt{P_i}\sum_{\substack{j = 1,j\neq i}}^N \frac{u_j v_j}{N\,\lambda_{\ell b}} \right|\right]\notag\\
&=E\left[\left|\sum_{\substack{j = 1}}^{N-1} \frac{u_j v_j}{N\,\lambda_{\ell b}} \right|\right]E\left[\left|\sum_{\substack{i \in\mathcal{T}, i\neq 0}}r_i^{-\frac{\alpha}{2}} \sqrt{P_i}\right|\right]\,,\label{Eqn:ExpectationID}
\end{align}
where $u_i$ and $v_i$ are i.i.d. $\mathcal{CN}(0,1)$ random variables. Note that the expectation in \eqref{Eqn:ExpectationID} factors due to the isotropic nature of Gaussian random vectors. Note that one of the standard methods to prove the weak law of large numbers is to show that the sample mean of zero-mean i.i.d. random variables converges in the mean-square sense to zero. Since convergence in mean square implies convergence in mean,
the first expectation on the RHS of \eqref{Eqn:ExpectationID} converges to zero.  Since the mobiles which contribute towards the pilot
contamination are located outside the cell at the origin, for $i \neq 0, i \in \mathcal{T}$,  $r_i > D_{\text{min}}$, where $D_{\text{min}}$ is the radius of the largest circle which is wholly contained in the cell at the origin.  Since $\alpha > 4$, from   Section III.A of \cite{HaenggiJSAC} , $E\left[\left|\sum_{\substack{i \in\mathcal{T}, i\neq 1}}r_i^{-{\alpha/2}} \right|\right]$ is bounded. Thus, we have \eqref{Eqn:ExpectationID} $\to 0$,
which implies that the first term in the absolute value in $\eqref{Eqn:SINRExpand}$ converges in probability to zero. From Theorem 1, the second term in the absolute value in $\eqref{Eqn:SINRExpand}$ converges in probability to ${r_0^{-\alpha}\beta}$. Thus, the following holds in probability
\begin{align}
N^{-\alpha}S\to r_0^{-2\alpha}\beta^2. \label{Eqn:ContaminatedSignalPowerConv}
\end{align}
Next,  consider the interference power normalized by $N^{-\alpha/2}$
\begin{align}
&N^{-\frac{\alpha}{2}}I= N^{-\frac{\alpha}{2}}\mathbf{w}^\dagger\sum_{i = 1}^n  r_i^{-\alpha}P_i\mathbf{g}_i\mathbf{g}_i^\dagger\mathbf{w} \nonumber\\
&=\frac1N\hat{\mathbf{h}}^\dagger  \left(\sum_{i = 1}^{n} \frac1N p_i \mathbf{g}_i\mathbf{g}_i^\dagger\right)^{-1}\hat{\mathbf{h}}\label{Eqn:MeanInterfExpanded}\,.\\
&= \frac1N \sum_{j\in\mathcal{T}}r_j^{-\frac{\alpha}{2}}\sqrt{P_j}\! \sum_{k\in\mathcal{T}} r_k^{-\frac{\alpha}{2}}\sqrt{P_k}\mathbf{g}_j^\dagger\left(\sum_{i = 1}^{n} \frac1Np_i \mathbf{g}_i\mathbf{g}_i^\dagger\right)^{-1}\mathbf{g}_k \,.\nonumber \\
&=  \sum_{j\in\mathcal{T}}r_j^{-\frac{\alpha}{2}}\sqrt{P_j}\!\! \sum_{\substack{k\in\mathcal{T},k \neq j}} r_k^{-\frac{\alpha}{2}}\sqrt{P_k}\frac1N\mathbf{g}_j^\dagger\left(\sum_{i = 1}^{n} \frac1Np_i \mathbf{g}_i\mathbf{g}_i^\dagger\right)^{-1}\!\!\!\!\mathbf{g}_k \nonumber \\
&+ \sum_{j\in\mathcal{T}}r_j^{-\alpha}P_j\frac1N\mathbf{g}_j^\dagger\left(\sum_{i = 1}^{n} \frac1Np_i \mathbf{g}_i\mathbf{g}_i^\dagger\right)^{-1}\mathbf{g}_j \,. \label{Eqn:FirstSigPow}
\end{align}
The first term on the RHS of \eqref{Eqn:FirstSigPow}, has a similar form to the first term in the absolute value in \eqref{Eqn:SINRExpand}, which was shown to converge in probability to zero. Following a similar sequence of steps, we can show that  the first term on the RHS of \eqref{Eqn:FirstSigPow} goes to zero as $n,N,R\to\infty$ which yields
\begin{align}
&\lim_{N\to\infty} N^{-\frac{\alpha}{2}}I =\lim_{N\to\infty}\frac1N\!\! \sum_{j\in\mathcal{T}}r_j^{-{\alpha}}{P_j}  \mathbf{g}_j^\dagger\!\!\left(\sum_{i = 1}^n \frac1N p_i  \mathbf{g}_i\mathbf{g}_i^\dagger\right)^{-1}\!\!\!\! \mathbf{g}_j\notag\\
&=\lim_{N\to\infty}\sum_{j\in\mathcal{T}, j \neq 0}  \frac{\frac1N r_j^{-{\alpha}}{P_j}\mathbf{g}_j^\dagger\left(\sum_{i = 1, i\neq j}^n\frac1N p_i  \mathbf{g}_i\mathbf{g}_i^\dagger\right)^{-1} \mathbf{g}_j}{1+ \frac1N p_j\mathbf{g}_j^\dagger\left(\sum_{i = 1, i\neq j}^n \frac1N p_i  \mathbf{g}_i\mathbf{g}_i^\dagger\right)^{-1} \mathbf{g}_j}\notag\\
&+ \frac1N r_0^{-{\alpha}}{P_j}  \mathbf{g}_0^\dagger\left(\sum_{i = 1}^n \frac1N p_i  \mathbf{g}_i\mathbf{g}_i^\dagger\right)^{-1} \mathbf{g}_0\notag\\
&= r_0^{-{\alpha}}{} \beta + \sum_{j\in\mathcal{T}, j\neq0}  \frac{r_j^{-\alpha}{P_j}\beta}{1+ \bar{p}_j\beta}, = \bar{\beta} \text{ in probability.}\;\;\;\;\;\;\;\;\;\;\;\;\;\;\;\\
&\leq r_0^{-{\alpha}}{} \beta + \sum_{j\in\mathcal{T}, j\neq0}  {r_j^{-\alpha}{P_j}\beta}\,,\;\;\;\;\;\;\;\;\;\;\;\;\;\;\;
\label{Eqn:InterfConverge}
\end{align}
where  second equality is from applying the Sherman-Morrison-Woodbury matrix inversion lemma. The theorem is proved by dividing \eqref{Eqn:ContaminatedSignalPowerConv} by \eqref{Eqn:InterfConverge}.

\bibliography{IEEEabrv,main}

\end{document}